\documentclass[aps,prb,reprint,superscriptaddress, floatfix]{revtex4-2}

\usepackage{graphicx, amsmath, amssymb}
\usepackage[version=4]{mhchem}
\usepackage{color}
\usepackage{lineno,hyperref}
\usepackage[number-unit-product=~, range-units=single, range-phrase=\text{--}, list-units=single, list-final-separator={\text{, and }}]{siunitx}
\DeclareSIUnit\rydberg{\text{Ry}}
\usepackage{physics}

\newcommand{\sub}[1]{\ensuremath{_{\text{#1}}}}
\newcommand{\degreeC}[1]{\num{#1}\unit{\degreeCelsius}}
\newcommand{\percentage}[1]{\num{#1}\unit{\percent}}

\begin{document}


\title{Quasi-2D Fermi surface of superconducting line-nodal metal \ce{CaSb2}}


\author{Atsutoshi~Ikeda}
\email{aikeda@umd.edu}

\author{Shanta~Ranjan~Saha}
\affiliation{Maryland Quantum Materials Center and Department of Physics, University of Maryland, College Park MD 20742, USA}

\author{David~Graf}
\affiliation{National High Magnetic Field Laboratory, Tallahassee FL 32310, USA}

\author{Prathum~Saraf}
\author{Danila~Sergeevich~Sokratov}
\affiliation{Maryland Quantum Materials Center and Department of Physics, University of Maryland, College Park MD 20742, USA}

\author{Yajian~Hu}
\author{Hidemitsu~Takahashi}
\author{Soichiro~Yamane}
\affiliation{Departmemt of Physics, Kyoto University, Kyoto 606-8502, Japan}

\author{Anooja~Jayaraj}
\affiliation{Department of Physics, University of North Texas, Denton TX 76203, USA}

\author{Jagoda~S\l{}awi\'{n}ska}
\affiliation{Zernike Institute for Advanced Materials, University of Groningen, Groningen 9747AG, The Netherlands}

\author{Marco~Buongiorno~Nardelli}
\affiliation{Department of Physics, University of North Texas, Denton TX 76203, USA}
\affiliation{Santa Fe Institute, Santa Fe NM 87501, USA}

\author{Shingo~Yonezawa}
\author{Yoshiteru~Maeno}
\thanks{Present address: Toyota Riken--Kyoto University Research Center, Kyoto 606-8501, Japan}
\affiliation{Departmemt of Physics, Kyoto University, Kyoto 606-8502, Japan}

\author{Johnpierre~Paglione}
\email{paglione@umd.edu}
\affiliation{Maryland Quantum Materials Center and Department of Physics, University of Maryland, College Park MD 20742, USA}


\date{\today}

\begin{abstract}
We report on the Fermi surfaces and superconducting parameters of \ce{CaSb2} single crystals 
(superconducting below $T\sub{c}\sim\SI{1.8}{\kelvin}$) grown by the self-flux method.
The frequency of de-Haas--van-Alphen and Shubnikov--de-Haas oscillations evidences a quasi-two-dimensional (quasi-2D) Fermi surface,
consistent with one of the Fermi surfaces forming Dirac lines predicted by first-principles calculations.
Measurements in the superconducting state reveal that \ce{CaSb2} is close to a type-I superconductor 
with the Ginzburg--Landau (GL) parameter of around unity.
The temperature dependence of the upper critical field $H\sub{c2}$ is well described by a model considering two superconducting bands, 
and the enhancement of the effective mass estimated from $H\sub{c2}(\SI{0}{\kelvin})$ is consistent with the quasi-2D band observed by the quantum oscillations.
Our results indicate that a quasi-2D band forming Dirac lines contributes to the superconductivity in \ce{CaSb2}.
\end{abstract}


\maketitle


\section{Introduction\label{introduction}}


%



Topological superconductivity is often found in the vicinity of a topological normal state.
Indeed, the most promising candidate for the topological superconductivity 
is a doped topological insulator~\cite{Hor2010CuxBi2Se3SC, Fu2010CuxBi2Se3, Sasaki2011CuxBi2Se3, Matano2016CuxBi2Se3, Tajiri2017CuxBi2Se3}.
Superconductivity found in Dirac semimetals is also a candidate 
of topological superconductivity~\cite{Kobayashi2015DiracSC, Aggarwal2016Cd3As2, He2016Cd3As2, Oudah2016superconductivity}.
Furthermore, recent theoretical and experimental works revealed that the surface superconductivity can be topological even though the bulk superconductivity 
is trivial~\cite{Xu2016surfaceTCSinFeBasedSCtheory, Zhang2018TCSonSurfaceFeTeSe}.
It is important to investigate materials with a novel electronic band structure as a candidate host of a novel superconducting state.

Line-nodal semimetals are a new type of topological material, in which a band crossing is protected by the crystalline symmetry 
and remains gapless along a certain line~\cite{Fang2015NodalLine, Fang2016reviewTNL}.
In nodal-loop materials with the mirror symmetry, topological crystalline superconductivity and second-order topological superconductivity 
have been theoretically predicted~\cite{Shaourian2018NodalLoopSC}.
Although many materials are predicted to have nodal lines in the absence of spin--orbit coupling (SOC)~\cite{Yu2015TNLAntiperovskite, Yamakage2016CaAgX},
those nodal lines are in most cases gapped out by SOC to make the material a topological insulator or a point-nodal semimetal~\cite{Kobayashi2017NodalLine}.
To realize a nodal line robust against SOC, an additional symmetry such as nonsymmorphic symmetry is required~\cite{Fang2015NodalLine}.
Therefore, line-nodal materials with nonsymmorphic symmetry are a good platform to investigate 
the interplay between nodal lines and superconductivity on~\cite{Kobayashi2016NonsymmorphicSC}.

\begin{figure*}
\includegraphics[width=0.9\linewidth]{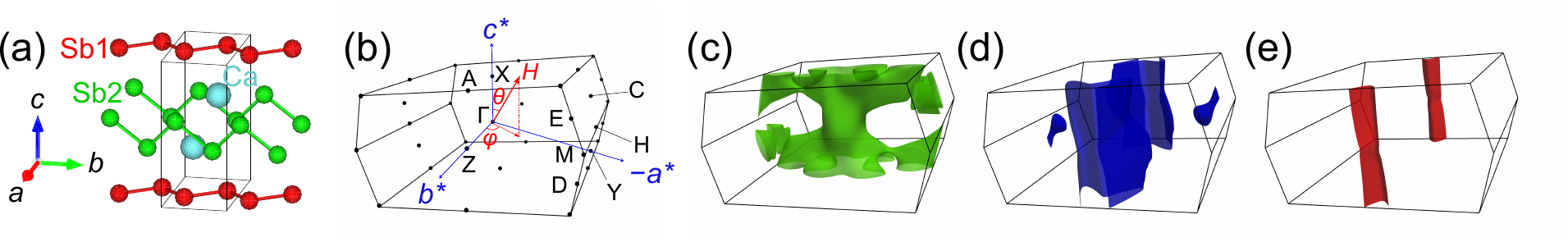}
\caption{(a)~Crystal structure of \ce{CaSb2}.
Light blue, red, and green spheres represent Ca, Sb1, and Sb2 atoms, respectively.
Chains of Sb1 are separated by Ca and Sb2.
The structure has a two-fold screw axis in the $b$ direction.
The figure was prepared with the program VESTA~\cite{Momma2011VESTA}.
(b)~Brillouin zone of \ce{CaSb2}.
The $z$ direction is chosen to be along the screw axis.
The field direction is specified by the polar angle $\theta$ from the $c^*$ axis and the azimuthal angle $\varphi$ from the $b^*$ axis.
Calculated Fermi surfaces of bands (c)~A, (d)~B, and (e)~C\@.
Bands B and C host Dirac lines at the Brillouin-zone boundary perpendicular to the $b^*$ axis.
The images of the Fermi surfaces were generated using ParaView 5.10.1~\cite{Ayachit2015ParaView}.}
\label{fig: Fermi surface}
\end{figure*}

\ce{CaSb2} is a candidate of the Dirac-line material whose nodes are protected by the screw symmetry 
of the nonsymmorphic crystalline structure belonging to the space group $P2_1/m$ (No.~11, $C_{2h}^2$)~\cite{Funada2019CaSb2}.
First-principles calculation shows that \ce{CaSb2} possesses three Fermi surfaces as shown in Fig.~\ref{fig: Fermi surface}.
Bands B and C, whose Fermi surfaces are nearly two-dimensional (2D), are degenerate along lines in the Brillouin-zone boundary, 
and those line degeneracies, or Dirac lines, cross the Fermi energy $E\sub{F}$.
\ce{CaSb2} also has a topologically trivial three-dimensional (3D) Fermi surface from band A around the $\Gamma$ point of the Brillouin zone.

Recently, some of the present authors discovered superconductivity in \ce{CaSb2} 
with a transition temperature of $T\sub{c}\sim\SI{1.7}{\kelvin}$~\cite{Ikeda2020CaSb2}.
While the measurements of the nuclear quadrupole resonance suggest a conventional superconductivity~\cite{Takahashi2021CaSb2NQR}, 
an unusual peak in $T\sub{c}$ under hydrostatic pressure implying a complex superconducting state 
was reported on a polycrystalline sample~\cite{Kitagawa2021CaSb2pressure}.
Furthermore, a theoretical analysis based on the symmetry indicators revealed that \ce{CaSb2} can exhibit 
line-nodal superconductivity reflecting the band structure in the normal state~\cite{Ono2021CaSb2SymmetryIndicator}.
Although there is a report of de-Haas--van-Alphen (dHvA) oscillations with the field in the $c^*$ direction~\cite{Oudah2022CaSb2Crystal}, the authors 
of this work did not find a quantitative match between theory and experiment possibly because ``the as-grown single crystals are self-doped.''
Moreover, the 3D shapes of the Fermi surfaces based on the field-angle dependence of the quantum oscillations were not obtained.
To fully understand the possible interplay between the superconductivity and Dirac lines, 
anisotropy of the superconducting state and underlying normal-state band structure (fermiology) is important.

In this paper, we report on the field-angle dependence of quantum oscillations and 
superconducting parameters of \ce{CaSb2} single crystals for the first time.
The field-angle dependence of the frequency of dHvA and Shubnikov--de-Haas (SdH) oscillations indicates a quasi-2D Fermi surface.
This Fermi surface hosts Dirac lines according to first-principles calculations, 
and therefore our observation provides indirect evidence of the Dirac lines crossing $E\sub{F}$.
Via resistance and magnetization with $H\parallel c^*$ and $H\perp c^*$, we estimated the anisotropy in the Ginzburg--Landau (GL) coherence length $\xi$, 
the GL parameter $\kappa$, and the penetration depth $\lambda$.
We found that the quasi-2D Fermi surface observed by the quantum oscillations is likely playing a role in the superconductivity in \ce{CaSb2}.

\section{Experiment\label{experiment}}
Single crystals of \ce{CaSb2} were grown by a self-flux method.
Ca (Sigma-Aldrich, \percentage{99.99}) and Sb (Alfa Aesar, \percentage{99.9999}) were placed in an alumina crucible (LSP ceramics) with a molar ratio of 1:5,
and the crucible was sealed in a quartz tube under \SI{0.3}{atm} of argon at room temperature.
The tube was heated in a box furnace (MTI Corporation, KSL-1100X) up to \degreeC{1000} in \SI{12}{\hour}, kept at that temperature for \SI{6}{\hour}, 
cooled down to \degreeC{740} in \SI{14}{\hour}, and then cooled down slowly to \degreeC{610} at a rate of \num{-1}\unit[per-mode=symbol]{\degreeCelsius\per\hour}.
The remaining flux was removed by centrifugation.

The material and orientation of crystals were identified with an x-ray diffractometer (Rigaku, MiniFlex600) using the Cu-$K\alpha$ radiation.
Crystals of \ce{CaSb2} are platelets with a thickness of about \SI{80}{\micro\meter}, 
and we confirmed that the crystal grows in the $ab$ plane as reported in Supplemental Material~\cite{Supplement}.

Direct-current (DC) magnetization $M$ above \SI{1.8}{\kelvin} was measured with a commercial magnetometer (Quantum Design, MPMS3).
DC magnetization down to \SI{0.5}{\kelvin} was measured with a magnetometer (Quantum Design, MPMS-XL) equipped with a \ce{^3He} refrigerator (IQUANTUM, iHelium3).
Alternating-current (AC) magnetic susceptibility was measured with a lock-in amplifier (Stanford Research Systems, SR830) 
using a homemade susceptometer similar to that reported in Ref.~\cite{Yonezawa2015ADR} compatible 
with the adiabatic-demagnetization-refrigerator (ADR) option of commercial equipment (Quantum Design, PPMS)\@.
Our crystal exhibits a single transition at \SI{1.8}{\kelvin}~\cite{Supplement}, while a polycrystalline sample 
used in the original report~\cite{Ikeda2020CaSb2} showed double transitions originating from the bulk and the grain boundary.

Magnetic torque $\tau$ was measured at the National High Magnetic Field Laboratory 
(Tallahassee, Florida, USA) under the DC field up to $\mu_0H=\SI{35}{\tesla}$ ($\mu_0$ is the magnetic permeability of vacuum).
We measured two samples A and B\@.
The samples were mounted on self-sensing cantilevers with a length of \SI{300}{\micro\meter} (SCL-Sensor Tech, Vienna, Austria),
and the cantilevers were placed in a \ce{^3He} refrigerator.
Piezo-torque magnetometry was performed with a balanced Wheatstone bridge that uses two piezoresistive paths 
on the cantilever (with and without the samples) as well as two resistors at room temperature 
that can be adjusted to balance the circuit.
The voltage across the Wheatstone bridge was measured using a lock-in amplifier (Stanford Research Systems, SR860).

Electrical resistance $R$ was measured by a standard four-probe method in PPMS\@.
Gold wires with a diameter of \SI{25}{\micro\meter} (California Fine Wires, \percentage{99.99}) 
were attached to the crystal using silver epoxy (Epoxy Technology, H20E)\@.
The current was in the $ab$ plane, but the in-plane angle is unknown.
The horizontal rotator option was used for angular dependence of the SdH oscillation, 
and the ADR option was used for measurements below \SI{1.8}{\kelvin}.
Our crystal shows good metallic behavior, with a residual resistivity of \SI{0.77}{\micro\ohm.\centi\meter} and a residual-resistance ratio (RRR) of 184.
The magnetoresistance reaches \percentage{6000} with an out-of-plane field of $\mu_0H=\SI{14}{\tesla}$,
while the magnetoresistance is \percentage{370} with in-plane field~\cite{Supplement}.
Large RRR and magnetoresistance ensure the good quality of our crystals.

\begin{figure*}
\includegraphics[width=0.9\linewidth]{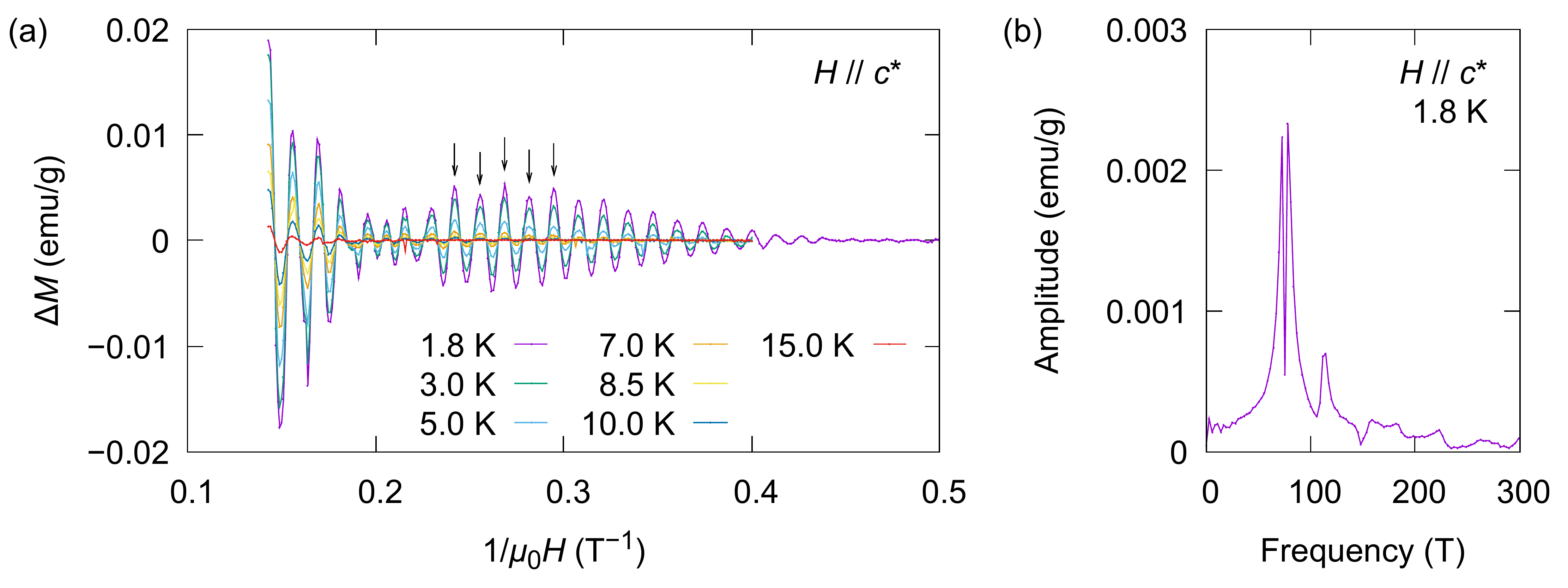}
\caption{(a)~de-Haas--van-Alphen (dHvA) oscillations of \ce{CaSb2} observed in DC magnetization under the field in the $c^*$ direction.
The beating indicates the presence of two frequencies: \qtylist{73.76; 77.98}{\tesla} according to the fitting with the Lifshitz--Kosevich formula.
In addition, the staggered amplitude indicated by arrows suggests another frequency of \SI{112.55}{\tesla}.
(b)~Fourier transform of the dHvA oscillations.
The resolution of the frequency is \SI{2.8}{\tesla}.
There are three distinct peaks at \qtylist{73; 78; 114}{\tesla}.}
\label{fig: dHvA}
\end{figure*}

First-principles calculations based on the density functional theory (DFT) were performed using the QUANTUM ESPRESSO package~\cite{QE-2009, QE-2017}.
We considered fully relativistic pseudopotentials from the pslibrary1.0.0 database~\cite{Perdew1996PBE, Kresse1990pseudopotential}.
The ion--electron interaction was treated with the projector augmented wave pseudopotentials, 
and the exchange and correlation interaction was treated within the generalized gradient approximation~\cite{DalCorso2014pseudopotential}.
The electron wavefunctions were expanded in a plane-wave basis set with a cutoff of \SI{45}{\rydberg}.
The internal degrees of freedom were relaxed until the forces on each atom were less than \SI{e-3}{\rydberg/\bohr}, where \si{\bohr} is the Bohr radius.
The Brillouin-zone sampling for the DFT calculation was performed using the Monkhorst--Pack scheme with a $16\times30\times32$ $k$-point grid, 
and SOC was taken into account self-consistently.
The Fermi surfaces were calculated on a further interpolated mesh of $48\times90\times96$ using the PAOFLOW code~\cite{BuongiornoNardelli2018PAOFLOW, Cerasoli2021PAOFLOW2}.
The quantum-oscillation frequencies expected for the calculated Fermi surfaces were evaluated using the SKEAF code~\cite{Rourke2012SKEAF}.
The calculated band structure is presented in Supplemental Material~\cite{Supplement}.

\section{Results\label{results}}
\subsection{Normal state}
Firstly, we present the dHvA oscillations of \ce{CaSb2} measured using DC magnetization in Fig.~\ref{fig: dHvA}(a).
The magnetic field is along the $c^*$ axis.
(\ce{CaSb2} has a monoclinic structure with $\beta=\ang{106.3}$, 
and thus we use the $c^*$ direction to indicate that the field is perpendicular to the $ab$ plane.)
We observed clear oscillations down to \SI{2.3}{\tesla} at \SI{1.8}{\kelvin} or up to \SI{15}{\kelvin} at \SI{7}{\tesla}.
The oscillations exhibit a beating, evidencing the existence of two frequencies close to each other.
In addition, the staggered amplitude of the beat (indicated by arrows) suggests another frequency.
We comment that we did not observe dHvA oscillations with the in-plane field up to \SI{7}{\tesla}.
The Fourier transform of the dHvA oscillations at \SI{1.8}{\kelvin} is shown in Fig.~\ref{fig: dHvA}(b).
We see three distinct peaks;
the peaks at \SIlist{73; 78}{\tesla} correspond to the beating in the raw oscillations and the one at \SI{114}{\tesla} to the staggered amplitude.

\begin{table*}
\caption{Parameters characterizing the de-Haas--van-Alphen oscillations of \ce{CaSb2} at different polar angles $\theta$: 
frequency $F$, effective mass $m^*$ divided by the electron mass $m_e$, Dingle temperature $T\sub{D}$, 
	quantum mobility $\mu\sub{q}=e\hbar/(2\pi m^*k\sub{B}T\sub{D})$, and phase shift $\delta$.
Parameters at $\theta=\ang{0}$ were estimated from magnetization, 
	and those at \ang{-10} and \ang{40} were from magnetic torque of samples A and B, respectively.}
\label{tbl: dHvA}
\begin{ruledtabular}
	\begin{tabular}{rS[table-format=3.2, table-figures-uncertainty=1]S[table-format=1.4, table-figures-uncertainty=2]S[table-format=2.2, table-figures-uncertainty=2]S[table-format=1.3, table-figures-uncertainty=2]S[table-format=+1.3, table-figures-uncertainty=2]c}
		{$\theta$\:\:} & {$F$ (\unit{\tesla})} & {$m^*/m_e$} & {$T\sub{D}$ (\unit{\kelvin})} & {$\mu\sub{q}$ (\unit[per-mode=symbol]{\meter^2\volt^{-1}\second^{-1}})} & {$\delta$} & {Possible origin} \\
		\hline
		        & 73.76+-0.03 & 0.1619+-0.0012 & 3.35+-0.10 & 0.394+-0.014 & -0.429+-0.008 & Band C \\
		\ang{0} & 77.98+-0.03 & 0.1616+-0.0012 & 3.30+-0.09 & 0.400+-0.014 & 0.211+-0.007 & Band C \\
		        & 112.55+-0.09 & 0.154+-0.005 & 3.5+-0.3 & 0.40+-0.05 & 0.09+-0.02 & Band A (or B) \\
		\hline
			& 74.73+-0.04 & 0.1689+-0.0010 & 2.5+-0.2 & 0.50+-0.03 & -0.422+-0.003 & Band C \\
		\ang{-10} & 80.46+-0.03 & 0.1708+-0.0008 & 1.86+-0.14 & 0.67+-0.05 & -0.009+-0.003 & Band C \\
		          & 115.66+-0.05 & 0.158+-0.003 & 1.5+-0.2 & 0.91+-0.11 & 0.141+-0.003 & Band A (or B) \\
		\hline
		         & 103.25+-0.08 & 0.239+-0.004 & 4.2+-0.3 & 0.21+-0.02 & -0.200+-0.007 & Band C \\
		\ang{40} & 107.5+-0.6 & 0.151+-0.013 & 18+-2 & 0.07+-0.02 & 0.18+-0.02 & Band C \\
		         & 127.4+-0.2 & 0.195+-0.013 & 5.8+-0.7 & 0.19+-0.04 & -0.276+-0.014 & Band A or B
\end{tabular}
\end{ruledtabular}
\end{table*}

We fitted the dHvA oscillations in Fig.~\ref{fig: dHvA}(a) with the Lifshitz--Kosevich (LK) formula~\cite{Shoenberg1984QO} with three components:
\begin{align}
&\quad\; \Delta M(B, T) \nonumber\\
&= -\sum_{i=1}^3 A_i\sqrt{B} R\sub{T}^i R\sub{D}^i R\sub{S}^i \sin\qty(2\pi\qty(F_iB-1/2+\delta_i)), \\
R\sub{T}^i &= X_iT/\sinh(X_iT), \\
R\sub{D}^i &= \exp\qty(-X_iT\sub{D}^i), \\
R\sub{S}^i &= \cos(\pi gm^*_i/(2m_e)), \\
X_i &= 2\pi^2k\sub{B}m^*_i/(\hbar eB),
\end{align}
where $g$ is the electron $g$ factor, $m^*$ is the effective mass, $m_e$ is the electron mass, 
$k\sub{B}$ is the Boltzmann constant, $\hbar$ is the reduced Planck constant, $e$ is the elementary charge, 
and $B$, $T$, and $F$ denote the flux density, temperature, and frequency, respectively.
$A$ is the amplitude of each oscillation component.
The phase shift $\delta$ reflects the dimensionality and the Berry phase;
the trivial Berry phase results in $\delta=0$ and $\pm1/8$ in 2D and 3D materials, respectively,
while a nontrivial Berry phase as in Dirac systems leads to $\delta=1/2$ and $3/8$ or $5/8$.
The fitting results are summarized in Table~\ref{tbl: dHvA}.
According to our first-principles calculation, expected oscillation frequencies around \SI{100}{\tesla} are \SIlist{211.1; 315.1}{\tesla} from the Fermi surface of band A, 
\SI{75.5}{\tesla} from the small pocket around the Y point of band B, and \SIlist{75.1; 184.4}{\tesla} from the minimum and maximum cross-sectional area of band C, 
if the chemical potential is shifted by \SI{-0.045}{\electronvolt} (hole doping).
The quasi-2D Fermi surface around the ZA line of band B is expected to exhibit oscillation frequencies above \SI{1300}{\tesla}.
We consider that the two frequencies \SIlist{73.8; 78.0}{\tesla} contributing to the beating observed in the experiment originate from band C;
the \SI[number-unit-product=-]{184.4}{\tesla} oscillation from band C in the calculation seems to have a much lower frequency in the experiment, 
or in other words, the Fermi surface is more 2D without a large deformation.
The field-angle dependence discussed later shows that both frequencies follow quasi-2D behavior, and thus neither is attributable to the 3D small pocket of band B\@.
The origin of the oscillation with \SI{112.6}{\tesla} is not clear, possibly related to band A or B\@.
The observed values of $\delta$ indicate that band C may have a nontrivial Berry phase.
Since the Dirac nodal lines protected by the nonsymmorphic symmetry are predicted on bands B and C,
the observation of one of these Fermi surfaces can be indirect evidence of the Dirac lines crossing $E\sub{F}$.

\begin{figure*}
\includegraphics[width=0.9\linewidth]{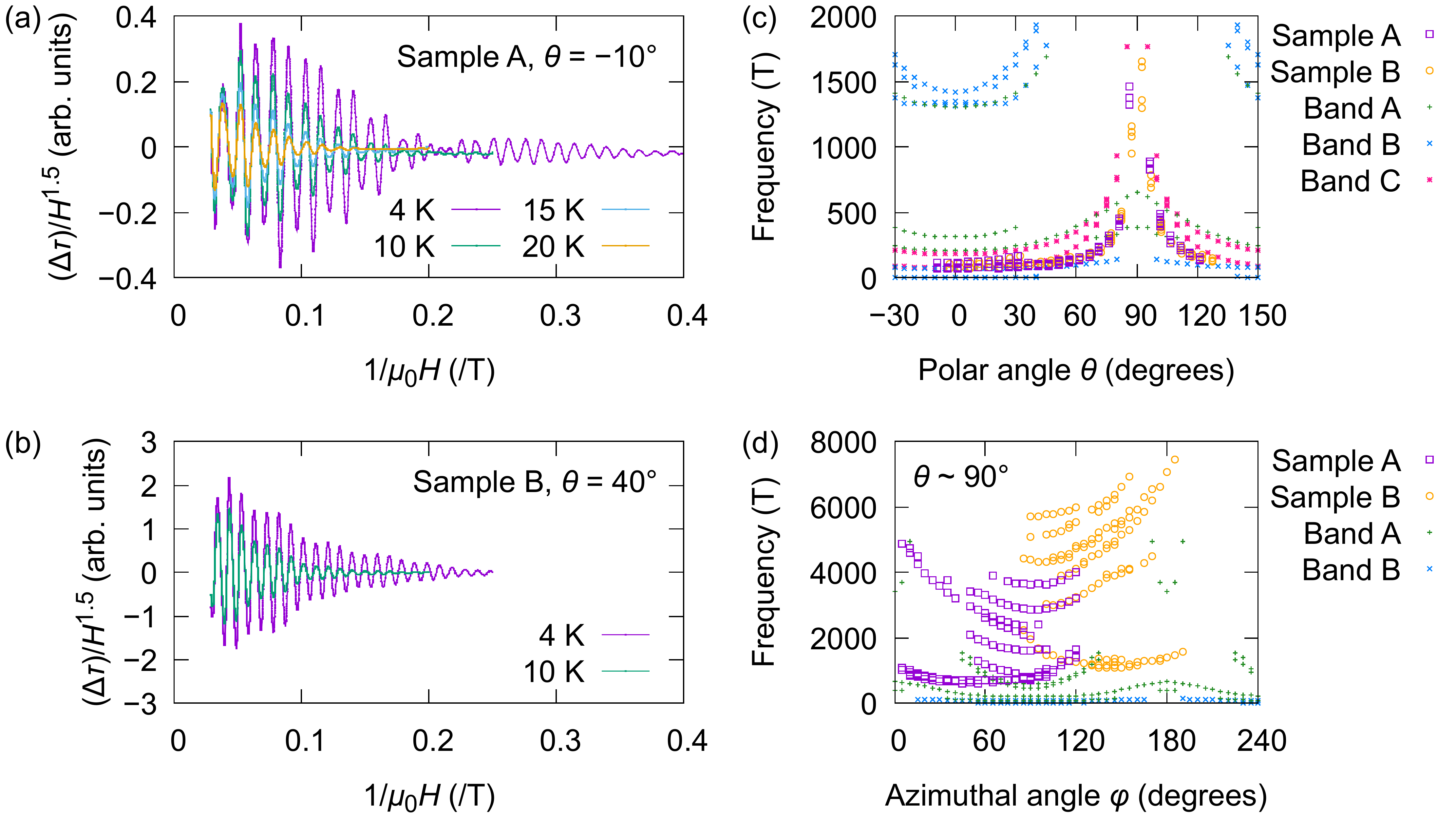}
	\caption{de-Haas--van-Alphen (dHvA) oscillations of \ce{CaSb2} measured by magnetic torque $\tau$.
	Temperature evolution of the oscillatory part of $\tau$ divided by magnetic field $H^{1.5}$ at (a)~$\theta=\ang{-10}$ and (b)~\ang{40}.
	As the field rotates from $c^*$ to the $ab$ plane, oscillation frequency becomes higher.
	(c)~$\theta$ dependence of the oscillation frequency along with the predictions at $\varphi=\ang{0}$ by the first-principles calculation.
	The divergence at $\theta=\ang{90}$ evidences that the observed oscillations originate from two-dimensional Fermi surfaces in \ce{CaSb2}.
	(d)~$\varphi$ dependence of the oscillation frequency at $\theta\sim\ang{90}$.
	The horizontal axis is calibrated so that the minimum frequency of each branch in the experiment appears at $\varphi\sim\ang{90}$ in accordance with the calculation.
	$\theta$ dependence was measured at $\varphi=``\ang{65}$'' for sample A and at ``\ang{235}'' for sample B, but these values may not reflect the crystallographic angle.
Detailed discussion of the $\varphi$ dependence is written in the main text.}
\label{fig: torque}
\end{figure*}

\begin{figure}[t!]
\includegraphics[width=0.85\linewidth]{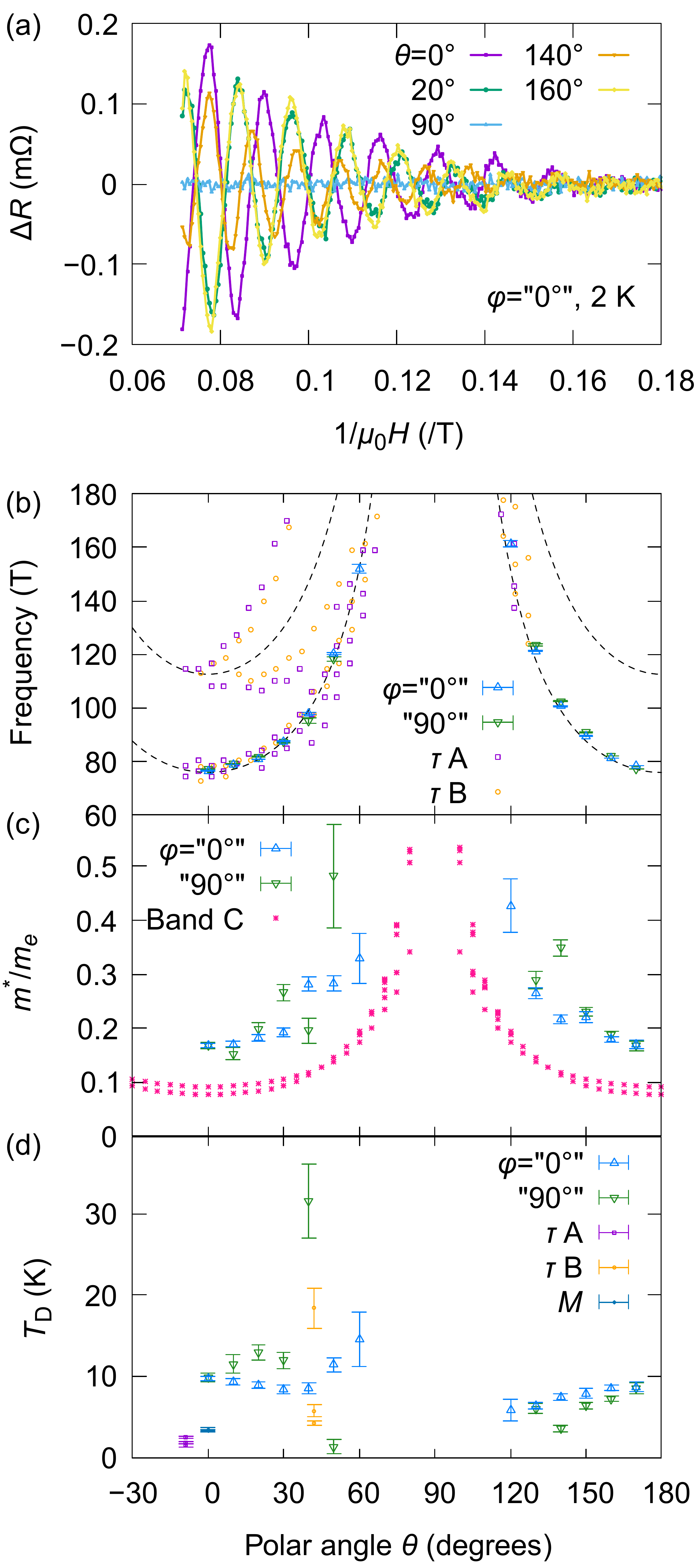}
\caption{Shubnikov--de-Haas (SdH) oscillations of \ce{CaSb2}.
	(a)~Oscillatory part of resistance at various field angles.
We observed oscillations with a frequency of \SI{76.54(6)}{\tesla} at $\theta=\ang{0}$.
	Data at \ang{20} and \ang{160} are on top of each other.
	Polar angle $\theta$ dependence of (b)~the SdH-oscillation frequency, (c)~the effective mass $m^*$ divided by the electron mass $m_e$, 
	and (d)~the Dingle temperature $T\sub{D}$, plotted with the data obtained from the torque samples A and B, the magnetization sample, 
and the first-principles calculation.
The dashed curves in (b) represent $A/\cos\theta$ with $A=\SI{75.87}{\tesla}$ and \SI{112.55}{\tesla} as determined from the magnetization at $\theta=\ang0$.
	The origin of the in-plane angle $\varphi$ is arbitrary.}
\label{fig: SdH}
\end{figure}

Comparing our results with the report in Ref.~\cite{Oudah2022CaSb2Crystal}, our crystal seems less hole-doped or closer to the stoichiometry.
The two frequencies originating from the electron Fermi surface of band C are larger by \qtyrange{0.5}{2.3}{\tesla}.
These shifts in the frequencies indicate more electron doping in our sample than that in Ref.~\cite{Oudah2022CaSb2Crystal}.
Assuming that the other frequency comes from the hole Fermi surface of band A, a smaller frequency also suggests a relative electron doping in our sample.
However, our first-principles calculation suggests that our crystal is hole-doped by \SI{0.045}{\electronvolt}.
Therefore, our crystal is less hole-doped rather than electron-doped.
The smaller doping may be reflected in the lower Dingle temperatures $T\sub{D}$ in our sample than in the previously reported crystal.

dHvA oscillations measured via magnetic torque are shown in Fig.~\ref{fig: torque}.
Figures~\ref{fig: torque}(a) and (b) show the raw oscillations at different field angles.
We see that the oscillation frequency becomes higher as the field rotates from the $c^*$ axis to the $ab$ plane.
The fitting results with the LK formula are summarized in Table~\ref{tbl: dHvA}.
The peak positions of the Fourier-transformed spectra are presented as a function of $\theta$ in Fig.~\ref{fig: torque}(c).
In the $\theta$ dependence, we see that the lower two frequencies diverge toward $\theta=\ang{90}$ following $1/\cos\theta$.
This divergence evidences the presence of a quasi-2D Fermi surface, consistent with band C\@.
The other frequency splits into two branches not following $1/\cos\theta$ as $\theta$ increases.
[A magnified view around low frequency is provided later as Fig.~\ref{fig: SdH}(b).]
These oscillations may be related to the \SI[number-unit-product=-]{211.1}{\tesla} oscillation from band A 
and the \SI[number-unit-product=-]{75.5}{\tesla} oscillation from band B\@.
The overall $\theta$ dependence evidences the presence of quasi-2D band C and other more 3D Fermi surfaces.

In the $\varphi$ dependence at $\theta\sim\ang{90}$ shown in Fig.~\ref{fig: torque}(d), we did not observe a good agreement between experiment and theory.
According to the first-principles calculation, we expect to see oscillations mainly from band A\@.
However, observed oscillations seem dominated by 2D Fermi surfaces for the following reason.
First, band C exhibits oscillations with about \SI{700}{\tesla} at around $\theta=\ang{80}$ and \SI{1100}{\tesla} at around \ang{85}.
Therefore, the main oscillations in Fig.~\ref{fig: torque}(d) (\SI{700}{\tesla} for sample A and \SI{1100}{\tesla} for sample B) 
are attributed to band C due to a small out-of-plane misalignment of the field.
Next, if we scale the frequencies for sample A by a factor of 1100/700, the data roughly match with those for sample B\@.
This fact suggests that all frequencies have the same $\theta$ dependence as the main oscillation from band C\@.
Thus, most of the oscillation frequencies observed in Fig.~\ref{fig: torque}(d) can be attributable to quasi-2D band C or the tubular part of band B\@.
We could not detect oscillations from band A probably because of a large scattering.
Since our calculation predicts that the effective mass for band A at $\theta=\ang{90}$ is reasonably small ($m^*/m_e\sim\numrange{0.1}{2.4}$ ), 
the dominant reason for the absence of quantum oscillations from band A will be a large scattering or high Dingle temperature.
As well as applying a higher field, eliminating the oscillations from bands B and C 
by precisely aligning the field direction will be necessary for the investigation of band A\@.

The SdH oscillations of \ce{CaSb2} at \SI{2}{\kelvin} are shown in Fig.~\ref{fig: SdH}(a).
We observed only one frequency component although there are multiple components in dHvA oscillations.
Probably the lower two frequencies observed in dHvA oscillations are not resolved,
and the other oscillation was not observed because of its small amplitude.
By fitting the data with the LK formula~\cite{Shoenberg1984QO, Ando1982SdHIn2D}, we extracted the frequency of the SdH oscillation, 
effective mass, and Dingle temperature at each angle as shown in Figs.~\ref{fig: SdH}(b--d).
Here, the origin of $\varphi$ is arbitrary since we do not know the in-plane direction of the crystal.
The frequency of the SdH oscillation matches with the lower two frequencies of the dHvA oscillations, diverging toward $\theta=\ang{90}$.
The effective mass is enhanced as the field rotates to the $ab$ plane, and the masses at \ang{60} and \ang{120} are nearly double the value at \ang{0}.
This increase in mass is consistent with the expected behavior for band C\@.
These angular dependences of the frequency and effective mass again confirm the presence of a quasi-2D Fermi surface in \ce{CaSb2}.
The Dingle temperature is almost independent of the angle and stays at around \SI{10}{\kelvin}.
The cause of higher Dingle temperature in the SdH oscillation than in the dHvA oscillation is not known;
possibly the crystal slightly degraded while curing Ag epoxy at \degreeC{150} or during repeated thermal cycles between room temperature and \SI{2}{\kelvin}.

\subsection{Superconducting state}
\begin{figure}
\includegraphics[width=0.85\linewidth]{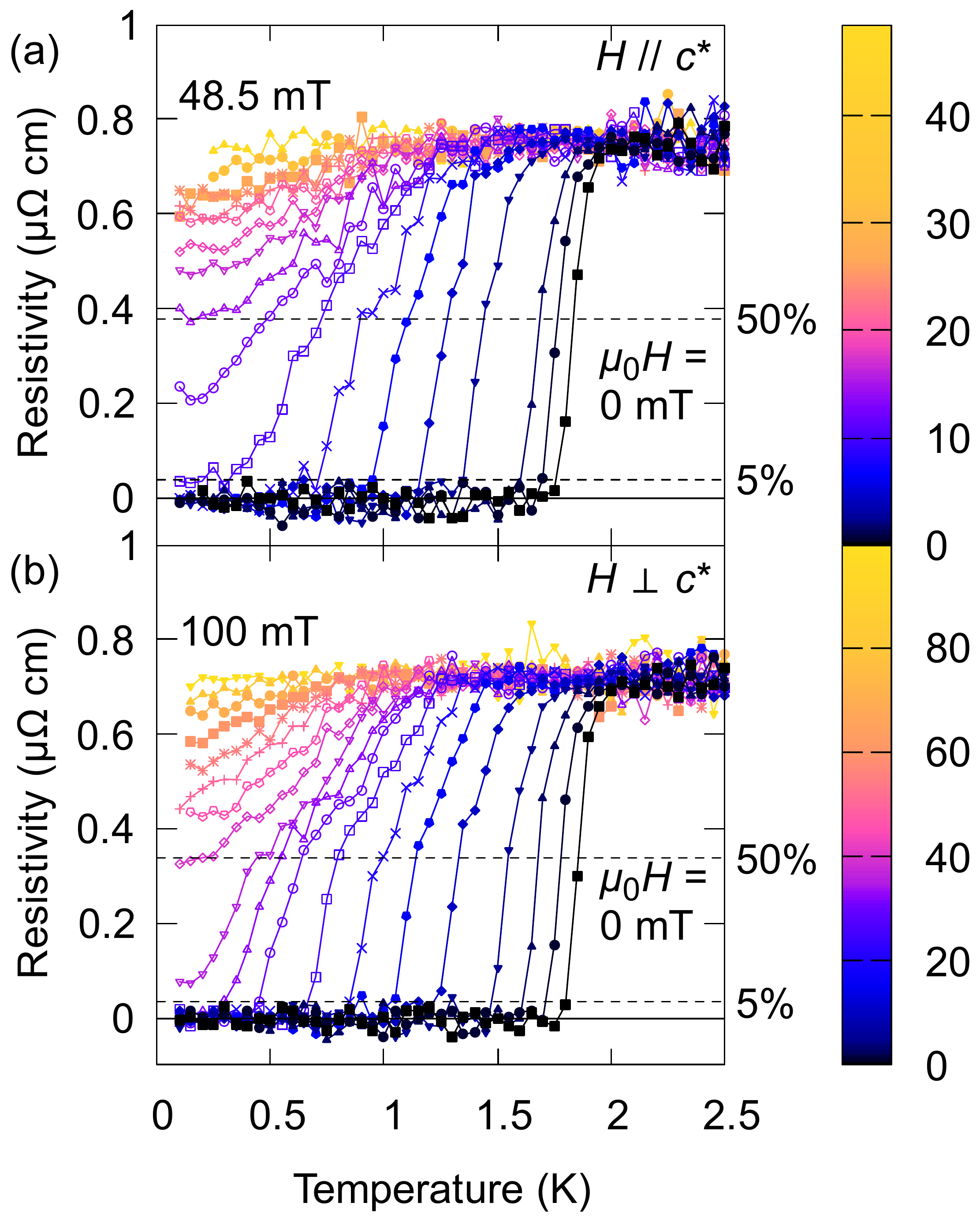}
\caption{Temperature dependence of the resistivity of \ce{CaSb2} with (a)~out-of-plane and (b)~in-plane fields.
Superconductivity is completely suppressed with $\mu_0H=\SI{48.5}{\milli\tesla} \parallel c^*$ and $\SI{100}{\milli\tesla} \perp c^*$.}
\label{fig: Hc2}
\end{figure}

We present in Fig.~\ref{fig: Hc2} the resistivity of \ce{CaSb2} below \SI{2.5}{\kelvin} under various field values.
Onset of the superconducting transition is suppressed to below \SI{0.1}{\kelvin} 
with $\mu_0H=\SI{48.5}{\milli\tesla} \parallel c^*$ [Fig.~\ref{fig: Hc2}(a)] and $\SI{100}{\milli\tesla} \perp c^*$ [Fig.~\ref{fig: Hc2}(b)].

\begin{figure}
\includegraphics[width=0.85\linewidth]{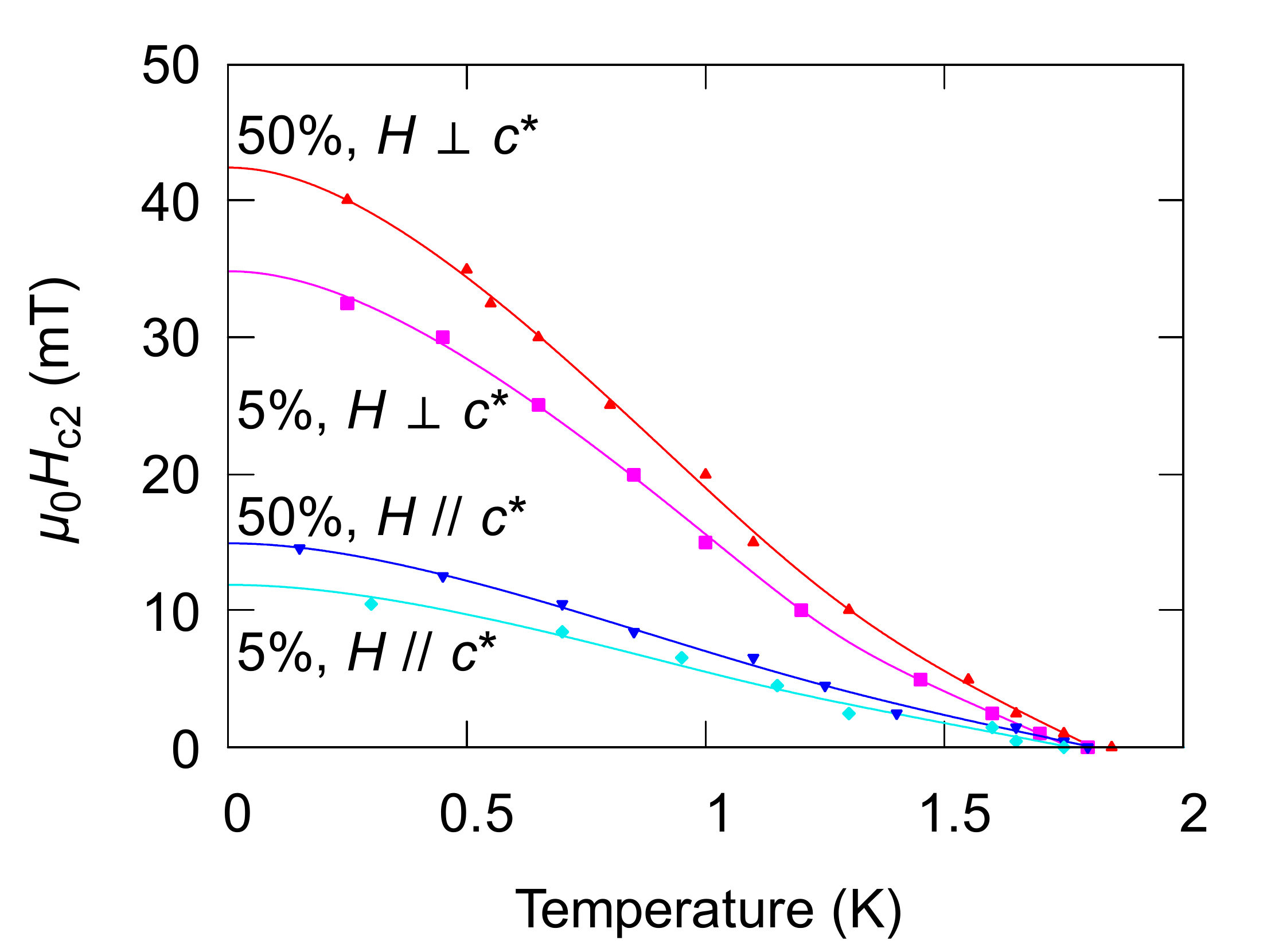}
\caption{Superconducting phase diagram of \ce{CaSb2}.
The curves represent a theoretical model for anisotropic two-gap superconductors.
The upper critical field at \SI{0}{\kelvin} is estimated to be $\mu_0H\sub{c2}=\qtyrange{12}{15}{\milli\tesla} \parallel c^*$ and $\qtyrange{35}{42}{\milli\tesla} \perp c^*$.}
\label{fig: phase diagram}
\end{figure}

Figure~\ref{fig: phase diagram} shows the superconducting phase diagram of \ce{CaSb2}.
We defined the transition temperature based on two criteria: \percentage{5} and \percentage{50} of the normal-state resistivity under a high magnetic field.
The temperature dependence of the upper critical field $H\sub{c2}$ cannot be fully described by the Werthamer--Helfand--Hohenberg relation~\cite{Werthamer1966Temperature} 
but can be well fitted by a theoretical model for anisotropic two-gap superconductors~\cite{Gurevich2003Hc2OfTwoGapSC}.
This nonmonotonic temperature dependence of the slope of the $T\sub{c}$--$H\sub{c2}$ curve was previously attributed to the effect of the grain boundary for a polycrystalline sample, 
but our study using a single crystal reveals that this behavior is intrinsic to \ce{CaSb2}.
Since \ce{CaSb2} has only one band (band A) which is not related to Dirac lines around $E\sub{F}$, the behavior of $H\sub{c2}$ consistent with multiple superconducting gaps 
indicates that at least one of the two bands forming Dirac lines (band B or C) contributes to the superconductivity of \ce{CaSb2}.

By fitting the data with the two-gap model, $H\sub{c2}(\SI{0}{\kelvin})$ was estimated to be $\mu_0H_{\textrm{c2} \parallel c^*}(\SI{0}{\kelvin})=\qtyrange{12}{15}{\milli\tesla}$ 
and $\mu_0H_{\textrm{c2} \perp c^*}(\SI{0}{\kelvin})=\qtyrange{35}{42}{\milli\tesla}$.
These values of $H\sub{c2}$ are about 10-times smaller than that of a polycrystalline sample~\cite{Ikeda2020CaSb2}.
We consider that a cleaner sample has led to a longer coherence length or smaller $H\sub{c2}$.
Compared to the report in Ref.~\cite{Oudah2022CaSb2Crystal}, $H_{\textrm{c2} \perp c^*}$ is about 1.5 times larger.
This difference may be due to the different measurement techniques or crystal quality, 
a $c^*$-axis component of the field by misalignment, or the anisotropy within the $ab$ plane.

The anisotropy of $H\sub{c2}(\SI{0}{\kelvin})$ can be explained by the field-angle dependence of the effective mass.
Considering that the coherence length in the GL theory at \SI{0}{\kelvin} is comparable to that in the Bardeen--Cooper--Schrieffer theory~\cite{Tinkham2004Superconductivity}:
$\xi\sub{GL}(\SI{0}{\kelvin}) = \sqrt{\Phi_0/(2\pi\mu_0H\sub{c2}(\SI{0}{\kelvin}))} \simeq \xi\sub{BCS} = \hbar v\sub{F}/(\pi\Delta)=\hbar^2 k\sub{F}/(m^*\pi\Delta)$ 
($\Phi_0$ is the flux quantum, $v\sub{F}$ and $k\sub{F}$ are the Fermi velocity and wavelength, respectively, and $\Delta$ is the superconducting gap), 
$(m^*_{H\perp c^*}/m^*_{H\parallel c^*})^2 \simeq H_{\textrm{c2} \perp c^*}/H_{\textrm{c2} \parallel c^*}=3$ suggests that $m^*$ is $\sqrt{3}$-times larger with $H\perp c^*$ than with $H\parallel c^*$.
This value is consistent with the fact that $m^*$ estimated from the SdH oscillation is nearly doubled at $\theta=\ang{60}$.
Therefore, the anisotropy of $H\sub{c2}$ can mainly be attributed to the field-angle dependence of $m^*$.
This fact indicates that the quasi-2D Fermi surface might be playing a major role near \SI{0}{\kelvin}.

\begin{figure}
\includegraphics[width=0.85\linewidth]{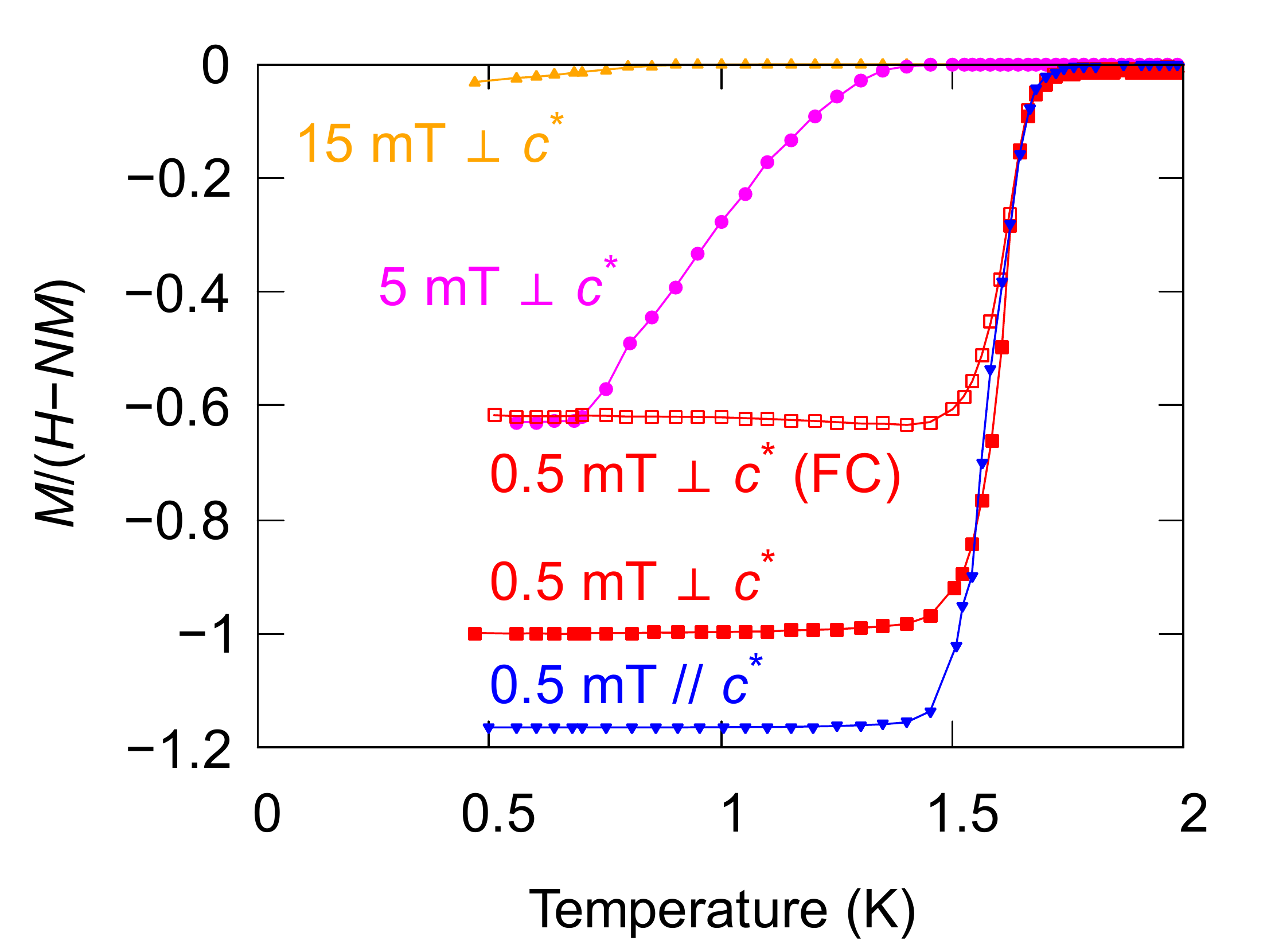}
\caption{Temperature dependence of DC magnetic susceptibility of \ce{CaSb2} after demagnetization correction.
The demagnetization factor was $N=0.53$ for $H\parallel c^*$ and $N=0.23$ for $H\perp c^*$.
	Data were obtained in the zero-field-cooling (ZFC) process except for the curve indicated as field cooling (FC).
	The shielding fraction under FC is \percentage{63} of that under ZFC, suggesting a weak pinning of the vortices.}
	\label{fig: M(T)}
\end{figure}

\begin{figure}
\includegraphics[width=0.85\linewidth]{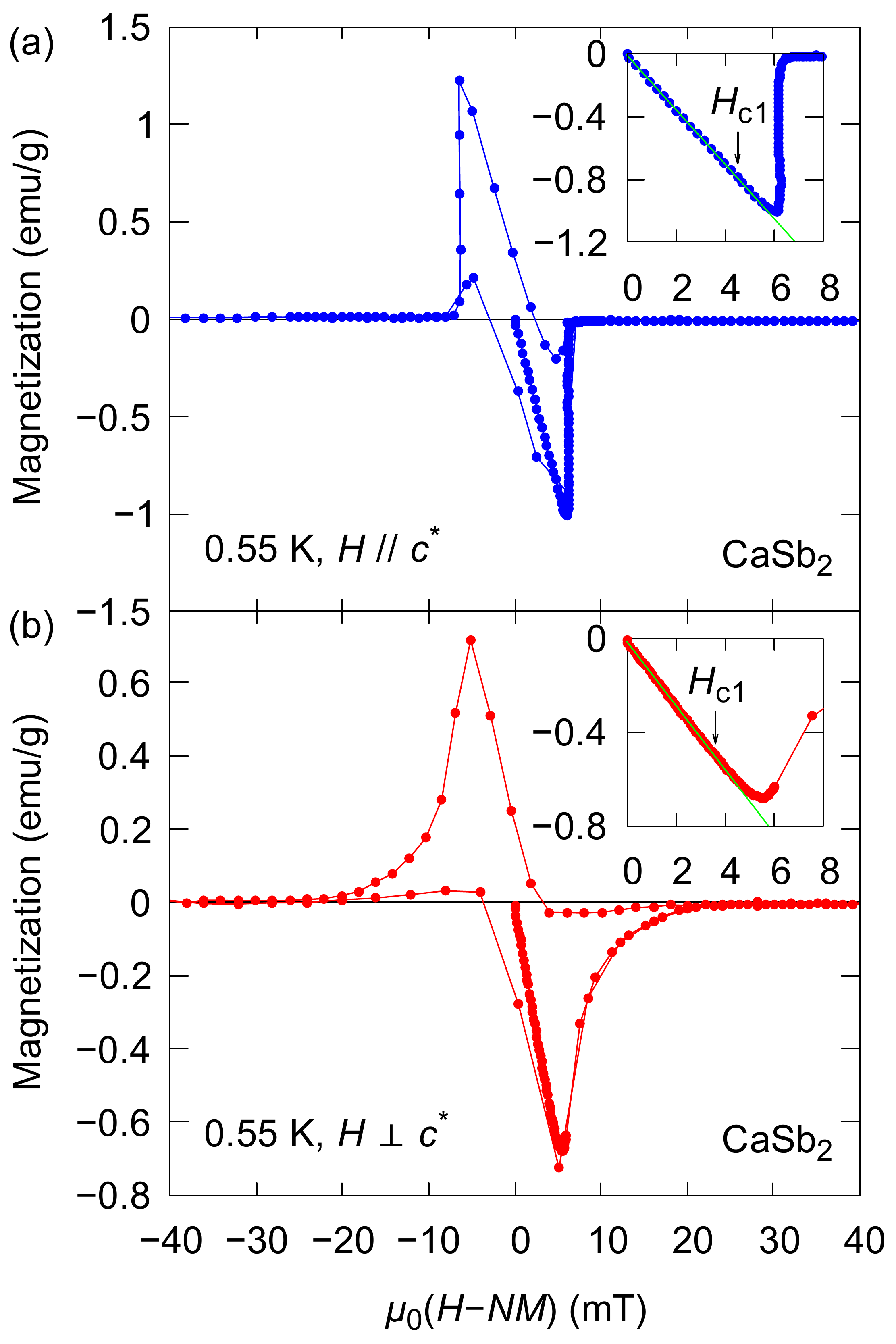}
\caption{Field dependence of DC magnetization of \ce{CaSb2} after demagnetization correction with (a)~out-of-plane and (b)~in-plane fields.
The demagnetization factor was $N=0.53$ for $H\parallel c^*$ and $N=0.23$ for $H\perp c^*$.
	We observed nearly type-I behavior for $H\parallel c^*$.
	Insets display enlarged views around the lower critical fields with linear fitting from $\mu_0H=\SI{0}{\milli\tesla}$.}
	\label{fig: M(H)}
\end{figure}

Figure~\ref{fig: M(T)} presents the temperature dependence of DC susceptibility after demagnetization correction.
Under both in-plane and out-of-plane fields, our sample exhibits strong diamagnetism of $M/H<-1$ before demagnetization correction.
The demagnetization factors were initially estimated by approximating the crystal by an ellipsoid, 
but this approximation led to a volume fraction larger than \percentage{100}, suggesting an underestimation of the demagnetization factors.
Thus, we adjusted the demagnetization factor for $H\perp c^*$ so that the volume fraction becomes \percentage{100} under \SI{0.5}{\milli\tesla} in the zero-field-cooling (ZFC) process.
For the demagnetization factor for $H\parallel c^*$, we used the maximum possible value that does not make the $M(H)$ curve shown in Fig.~\ref{fig: M(H)}(a) a multivalued function.
This value still leads to a volume fraction larger than \percentage{100}, but a further increase of the demagnetization factor leads to an unphysical $M(H)$ behavior.
With these demagnetization factors, the shielding fraction in the field-cooling (FC) process is \percentage{63} of that in the ZFC process.
Therefore, \ce{CaSb2} is a type-II superconductor with a weak vortex pinning.
The shielding fraction decreases with $\mu_0H=\SI{5}{\milli\tesla}$, 
indicating a lower critical field $H\sub{c1}$ of $\mu_0H\sub{c1}<\SI{5}{\milli\tesla}$.

Figure~\ref{fig: M(H)} shows the field dependence of magnetization under (a)~out-of-plane and (b)~in-plane fields
at \SI{0.55}{\kelvin} after demagnetization correction.
With $H\parallel c^*$, we observed nearly type-I behavior.
Superconducting parameters~\cite{Tinkham2004Superconductivity} estimated 
from Figs.~\ref{fig: phase diagram}, \ref{fig: M(T)}, and \ref{fig: M(H)} are summarized in Table~\ref{tbl: SC param}.
The GL coherence length was estimated as $\xi_{ab} = \sqrt{\Phi_0/(2\pi\mu_0H_{\textrm{c2} \parallel c^*})}$ and $\xi_{c^*} = \Phi_0/(2\pi\xi_{ab}\mu_0H_{\textrm{c2} \perp c^*})$.
$H\sub{c1}$ was estimated from the field range where the Pearson correlation coefficient between magnetization and field becomes maximum.
The thermodynamic critical field $H\sub{c}$ was estimated by integrating the $M(H)$ curve:
\begin{align}
\frac{\mu_0}{2}H\sub{c}^2 &= F\sub{n}-F\sub{s} \\
&= \frac{\mu_0}{2}\qty(\int_{H\sub{c2}}^0 M\dd{H} - \int_0^{-H\sub{c2}} M\dd{H}),
\end{align}
where $F\sub{n}$ and $F\sub{s}$ are the free energy of the normal and superconducting states, respectively.
Here, we averaged the two integrals of the $M(H)$ curve to cancel the effect from vortex pinning.
Both data with $H\parallel c^*$ and $H\perp c^*$ lead to a consistent value of $\mu_0H\sub{c}=\SI{6.0}{\milli\tesla}$.
Then, the GL parameter was calculated by $\kappa = H\sub{c2}/\qty(\sqrt{2}H\sub{c})$.
We comment that the commonly used relation for $\kappa\gg1/\sqrt{2}$~\cite{Abrikosov1988Metals}
\begin{align}
\frac{H\sub{c1}}{H\sub{c2}} &= \frac{\ln(\kappa+0.081)}{2\kappa^2}
\label{eqn: Hc1-Hc2}
\end{align}
is not valid for \ce{CaSb2}.
Indeed, Eq.~\ref{eqn: Hc1-Hc2} does not have a solution for $\kappa$ with the critical fields found in our measurements.
The penetration depth was estimated by $\lambda_{ab} = \kappa_{c^*}\xi_{ab}$ 
and $\lambda_{c^*} = \kappa_{ab}^2\xi_{c^*}/\kappa_{c^*}$ [or $\kappa_{ab}=\sqrt{\lambda_{ab}\lambda_{c^*}/\qty(\xi_{ab}\xi_{c^*})}$].
Our analysis resulted in a larger $\kappa$ than the previous report~\cite{Oudah2022CaSb2Crystal}, 
which uses Eq.~\ref{eqn: Hc1-Hc2} with a correction term.
We use a different estimation method taking the small value of $\kappa$ and the certain amount of anisotropy into account.
In any case, \ce{CaSb2} is still close to the type-I superconductor, especially with $H\parallel c^*$.

\begin{table}
\caption{Superconducting parameters of \ce{CaSb2}: in-plane and out-of-plane lower critical field $H\sub{c1}$, 
thermodynamic critical field $H\sub{c}$, upper critical field $H\sub{c2}$, Ginzburg--Landau (GL) coherence length $\xi$, 
penetration depth $\lambda$, and GL parameter $\kappa$.}
\label{tbl: SC param}
\begin{ruledtabular}
\begin{tabular}{ccc}
Direction                            & $ab$                            & $c^*$ \\
\hline
$\mu_0H\sub{c1}$(\SI{0.55}{\kelvin}) & \SI{3.6}{\milli\tesla}          & \SI{4.5}{\milli\tesla} \\
$\mu_0H\sub{c}$(\SI{0.55}{\kelvin})  & \multicolumn{2}{c}{\SI{6.0}{\milli\tesla}} \\
$\mu_0H\sub{c2}$(\SI{0}{\kelvin})    & \SIrange{35}{42}{\milli\tesla}  & \SIrange{12}{15}{\milli\tesla} \\
$\xi$(\SI{0}{\kelvin})               & \SIrange{149}{166}{\nano\meter} & \SIrange{52}{57}{\nano\meter} \\
$\lambda$(\SI{0}{\kelvin})           & \SIrange{182}{205}{\nano\meter} & \SIrange{530}{574}{\nano\meter} \\
$\kappa$                             & \numrange{3.2}{3.9}             & \numrange{1.1}{1.4}
\end{tabular}
\end{ruledtabular}
\end{table}

%


\section{Conclusion\label{conclusion}}
We synthesized single crystals of \ce{CaSb2} and observed a quasi-2D Fermi surface likely contributing to the superconductivity 
by the field-angle dependence of the quantum oscillations and superconducting parameters.
The field-angle dependence of the main dHvA- and SdH-oscillation frequency diverges following $1/\cos\theta$.
This result is consistent with one of the quasi-2D Fermi surfaces forming the Dirac nodal lines according to the first-principles calculation.
Therefore, our observation gives indirect evidence of the crossing of the Dirac lines across $E\sub{F}$ in \ce{CaSb2}.

In the superconducting state, the GL parameter is close to unity with $H\parallel c^*$.
Thus, \ce{CaSb2} is close to the type-I superconductor.
The temperature dependence of $H\sub{c2}$ is well described by the two-gap model, 
and thus at least one band forming Dirac lines is supposed to take part in superconductivity.
Our results, providing validity of the first-principles calculations and showing anisotropy of the superconducting parameters,
suggest that \ce{CaSb2} can be a good platform for investigations of the interplay between Dirac nodal lines in the normal state and superconductivity.


\begin{acknowledgments}
Research at the University of Maryland was supported by the US Department of Energy Award No.~DE-SC0019154 (magnetic and transport experiments), 
the Gordon and Betty Moore Foundation's EPiQS Initiative through Grant No.~GBMF9071 (materials synthesis), 
and the Maryland Quantum Materials Center.
The National High Magnetic Field Laboratory is supported by the National Science Foundation Cooperative Agreement No.~DMR-1644779 and the State of Florida.
DC magnetization down to \SI{0.5}{\kelvin} was measured in the Research Center for Low Temperature and Materials Sciences, Kyoto University.
Measurements in Kyoto were supported by the Japan Society for the Promotion of Science (JSPS) 
Core-to-Core Program (No.~JPJSCCA20170002) and the JSPS KAKENHI (No.~JP17H06136, JP20H05158, and JP20F20020).
Y.J.H. is supported by the JSPS Research Fellowship.
J.S. acknowledges the Rosalind Franklin Fellowship from the University of Groningen.
The theoretical calculations were carried out at the Texas Advanced Computing Center at the University of Texas, Austin.
M.B.N. and A.J. acknowledge support from the US Department of Energy through Grant No.~DE-SC0019432.
\end{acknowledgments}

\bibliography{antiperovskite}

\end{document}